\documentclass[%
 aip,
 amsmath,amssymb,
 reprint,%
]{revtex4-1}

\usepackage{graphicx}% Include figure files
\usepackage{dcolumn}% Align table columns on decimal point
\usepackage{bm}% bold math
\usepackage[utf8]{inputenc}
\usepackage[T1]{fontenc}
\usepackage{mathptmx}

\newcommand{\la}{\lambda}

\newcommand{\om}{\omega}

\newcommand{\ox}{\overline{x}}
\newcommand{\tk}{\widetilde{k}}
\newcommand{\tom}{\widetilde{\omega}}
\newcommand{\ttq}{\widetilde{q}}

\newcommand{\prt}{\partial}

\begin{document}

%\preprint{AIP/123-QED}

\title{Theory of quasi-simple dispersive shock waves and number of solitons evolved 
from a nonlinear pulse }

\author{ A. M. Kamchatnov}

\affiliation{Institute of Spectroscopy,
Russian Academy of Sciences, Troitsk, Moscow, 108840, Russia}
\altaffiliation[Also at ]{Moscow Institute of Physics and Technology, Institutsky
lane 9, Dolgoprudny, Moscow region, 141700, Russia.}%Lines break automatically or can be forced with \\

\email{kamch@isan.troitsk.ru.}

\date{\today}

\begin{abstract}
The theory of motion of edges of dispersive shock waves generated after wave breaking of simple waves
is developed. It is shown that this motion obeys Hamiltonian mechanics complemented by a Hopf-like 
equation for evolution of the background flow that interacts with edge wave packets or edge
solitons. A conjecture about existence of a certain symmetry between equations for the small-amplitude 
and soliton edges is formulated. In case of localized simple wave pulses propagating through a
quiescent medium this theory provided a new approach to derivation of an asymptotic formula for 
the number of solitons produced eventually from such a pulse.
\end{abstract}


% 47.35.Jk 	Wave breaking
% 02.30.Ik 	Integrable systems
% 02.30.Jr 	Partial differential equations
% 47.35.Fg 	Solitary waves

\maketitle

\begin{quotation}
In quite general situations, a localized intensive nonlinear wave pulse splits during its
evolution into two pulses propagating in opposite directions. Such individual pulses with
unidirectional propagation are called \textit{simple waves} and they can be described by
evolution of a single variable. Simple waves break with formation of dispersive shock waves
(DSWs) that can be represented as modulated nonlinear periodic waves whose evolution is 
governed in Gurevich-Pitaevskii approach by the Whitham modulation equations. We call such
type of DSWs \textit{quasi-simple shocks} and show that in this case the motion of the 
small-amplitude DSW edge is governed by the Hamilton equations with the dispersion law 
for linear waves playing the role of the Hamiltonian. The Hamilton equations
have an integral which plays the role of the limiting modulation parameter in the Whitham
system at this edge. On the basis of old Stokes' observation about expression of soliton's
speed in terms of the dispersion law of linear waves and other similar findings, we formulate 
a conjecture about relationship between limiting equations for the two edges. This theory
leads to derivation of an asymptotic formula for the number of solitons produced from
an initially localized simple wave pulse. The developed theory is applicable to a quite
wide class of nonlinear wave equations which is not limited to completely integrable equations.
\end{quotation}

\section{Introduction}\label{intro}

As is known, if a nonlinear wave system supports soliton-like propagation, then an intensive
enough initial pulse evolves eventually into a certain number $N$ of solitons and some
amount of linear radiation which is negligibly small for large $N$. Therefore, possibility
of prediction of this number $N$ for a given initial pulse is very important for the
theoretical description of behavior of nonlinear pulses in many experimental situations.
If the nonlinear wave equation is completely integrable, then this problem can be solved in
principle by considering the associated with this equation linear spectral problem: $N$ is
equal to the number of discrete eigenvalues for given initial data, whereas the eigenvalues
$\la_i$, $i=1,2,\ldots,N$, determine the parameters of solitons emerging from the pulse at
asymptotically large time $t\to\infty$ (see Ref.~\onlinecite{ggkm-67}). For large $N\gg1$ 
the quasi-classical method can be applied to the spectral problem which provides approximate 
values for $\la_i$ and simple asymptotic expression for $N$ (see Ref.~\onlinecite{karpman-67}). 
However, such a general method does not exist for non completely integrable equations. 
Nevertheless, some particular results can be obtained if we confine ourselves to the initial 
pulses of a simple-wave type and trace in some detail a gradual process of solitons formation 
from an initially smooth pulse. In fact, this restriction is not very strong since in 
hydrodynamic approximation with neglected dispersion effects any typical localized pulse 
splits during its evolution into two pulses propagating in opposite directions. If this 
splitting takes place at the stage of evolution before the wave breaking moment then the 
above condition is fulfilled due to the natural wave dynamics. We will consider in what
follows the simple-wave initial pulses only.

In dispersive nonlinear systems, wave breaking leads to
formation of a dispersive shock wave (DSW), that is a region of strong nonlinear 
oscillations. As was shown in Ref.~\onlinecite{gp-73}, such a region can be
presented as a modulated periodic solution of the wave equation under consideration and
then the Whitham modulation equations, Ref.~\onlinecite{whitham}, can be applied to description 
of its evolution. In Gurevich-Pitaevskii approach, a DSW degenerates at one its edge to a sequence 
of solitons and at another edge to a linear wave packet with vanishing amplitude. Each edge
propagates along the corresponding parts of the hydrodynamic simple-wave solution of
dispersionless approximation. Generally speaking, even evolution of initially simple-wave 
pulses can lead to formation of quite complicated wave structures with several DSWs,
rarefaction waves and plateau regions, if the system is not 
genuinely nonlinear, that is if its characteristic velocities can vanish at some values 
of wave amplitude, or the initial pulse profile has several local extrema or inflection points. 
However, if we confine ourselves to a simple-wave type of initial conditions with a single
local extremum of the amplitude for genuinely nonlinear systems, then a single DSW evolves 
after wave breaking moment. Situation simplifies even more, if the pulse propagates into
a quiescent medium. As was noticed in Ref.~\onlinecite{gkm-89} for the completely integrable
Korteweg-de Vries (KdV) equation, in this case the DSW is described by only two
varying parameters and it was called {\it quasi-simple} by analogy with hydrodynamical
simple waves with a single varying parameter. The shall generalize the notion of 
quasi-simple DSWs to all situations with wave-breaking of initially
simple wave smooth pulses. If such a DSW propagates through a quiescent medium, then 
this subclass of quasi-simple DSWs admits more complete investigations and even in this 
restricted formulation, the problem of description of DSW formation is applicable to a 
huge number of realistic experimental situations. In this paper we shall consider genuinely
nonlinear physical systems and simple-wave type of initial conditions for pulses
propagating into a quiescent medium.

\section{Formulation of the problem}

Here we define in more explicit terms the class of physical wave systems to which our approach 
can be applied.
We consider some nonlinear dispersive system and assume that in the so-called dispersionless
limit, when the higher order derivatives of physical variables are neglected, the
resulting equations can be written in a hydrodynamics-like form
\begin{equation}\label{eq8-1}
  \frac{\prt \rho}{\prt t}+\frac{\prt(\rho u)}{\prt x}=0,\quad
  \frac{\prt v}{\prt t}+v\frac{\prt v}{\prt x}+\frac{c^2}{\rho}\frac{\prt\rho}{\prt x}=0,
\end{equation}
where $\rho$ plays the role of ``density'', $v$ is the ``flow velocity'' and $c(\rho)$ has 
the meaning of the ``local sound velocity'' which is related with $\rho$ according to the
``equation of state'' $p=p(\rho)$ according to the relationship $c^2=dp/d\rho$. It is known that 
in many physical situations nonlinear wave equations can be written in this form (see, e.g.,
Refs.~\onlinecite{sg-69,gke-90}). The system (\ref{eq8-1}) has a standard for compressible fluid
dynamics form and can be cast to a diagonal Riemann form (see, e.g., Ref.~\onlinecite{LL-6})
\begin{equation}\label{eq8-2}
  \frac{\prt r_+}{\prt t}+v_+(r_+,r_-)\frac{\prt r_+}{\prt x}=0,\quad
  \frac{\prt r_-}{\prt t}+v_-(r_+,r_-)\frac{\prt r_-}{\prt x}=0,
\end{equation}
where
\begin{equation}\label{eq8-3}
  r_{\pm}=\frac{v}2\pm\frac12\int_{\rho_0}^{\rho}\frac{c(\rho)}{\rho}d\rho
\end{equation}
the Riemann velocities
\begin{equation}\label{eq8-4}
  v_{\pm}=v\pm c
\end{equation}
are expressed in terms of $r_{\pm}$ by means of solving Eqs.~(\ref{eq8-3}) with respect
to $v$ and $c=c(\rho)$ and substitution of the result into Eqs.~(\ref{eq8-4}).

In simple waves, one of the Riemann invariants $r_{\pm}$ is constant and for definiteness
we assume that this is $r_-$.  Besides that, we consider here pulses propagating into a
uniform quiescent medium with constant density $\rho=\rho_0$ and zero flow velocity
$v=0$, that is $r_-=0$ everywhere and $r_+=0$ outside the pulse. For such waves, the 
system (\ref{eq8-2}) reduces to the first equation only with $v_+=v_+(r_+,0)$,
$v_+(0,0)=c(\rho_0)$. Instead of $r_+$, we can choose for our convenience as a 
physical variable some other function $u=u(r_+)$ and change the reference frame by the 
replacement $x\to x+c(\rho_0)t$. Then the function $u(x,t)$ obeys the equation
\begin{equation}\label{eq8-5}
  \frac{\prt u}{\prt t}+V_0(u)\frac{\prt u}{\prt x}=0,
\end{equation}
where $V_0=v_+(r_+(u),0)$ and $V_0\to0$ for $u\to0$. As was conjectured by Gurevich
and Meshcherkin in Ref.~\onlinecite{gm-84}, the constant Riemann invariant $r_-$ 
preserves the same value at both edges of the DSW in spite of its fast
oscillations within the DSW region. In systems described by completely integrable 
equations, this condition is fulfilled by the Gurevich-Pitaevskii construction of 
the solution of Whitham's equations, and Gurevich and Meshcherkin generalized this 
property to non-completely integrable situations.

The full system of wave equations, which includes higher order derivatives of $\rho$ and $v$,
can be linearized with respect to small deviations $\rho'$, $v'$ from their ``background''
values $\rho,v$ which can be considered locally as constant. Then the linear wave solutions
$\rho',v'\propto\exp[i(kx-\om t)]$ yield two branches
\begin{equation}\label{eq8-6}
  \om=\om_{\pm}(\rho,v,k)=\om_{\pm}(r_+,r_-,k)
\end{equation}
of the dispersion law. Again we put here $r_-=0, r_+=r_+(u)$ and take the branch for
which the phase velocity $V_{\pm}=\om_{\pm}/k$ converges to $V_0(u)$ in the limit $k\to0$.
This means that we consider linear waves for which their long wavelength limit is
consistent with linearization of Eq.~(\ref{eq8-5}): dispersionless evolution coincides
locally with unidirectional propagation of long wavelength linear waves. As a result, 
we arrive at the dispersion law
\begin{equation}\label{eq9-7}
  \om=\om(u,k),\qquad V(u,k)={\om(u,k)}/k
\end{equation}
with its dispersionless limit
\begin{equation}\label{eq9-8}
  \om=V_0(u)k,\quad V_0(u)=V(u,0),\quad V_0(0)=0.
\end{equation}

\begin{figure}
\includegraphics[width=8.5cm]{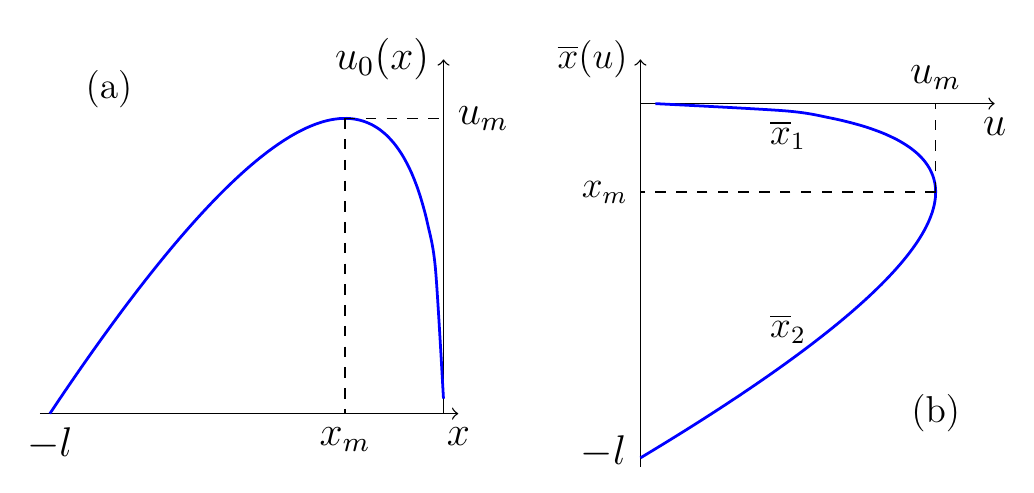}
\caption{(a) The initial profile $u_0(x)$. (b) Two
branches $\ox_1(u)$ and $\ox_2(u)$ of the
inverse function.}
\label{fig1}
\end{figure}

For definiteness, we will consider physical systems with $V_0(u)>0$ and negative dispersion
($d^2\om/dk^2<0$) which support ``bright'' soliton solutions in the form of humps
of the variable $u$ propagating along the background with $u=0$. We assume that the 
initial distribution $u_0(x)=u(x,0)$ is a smooth function and belongs to the
simple-wave type of unidirectional propagation. Its dispersionless evolution
according to Eq.~(\ref{eq8-5}) leads to steepening of the front so that the
wave breaks at some moment of time. To simplify the notation, we take $t=0$ as a 
wave-breaking moment and choose a localized form of the initial pulse with 
$u_0(x)>0$ for $-l<x<0$ and $u_0(x)=0$ outside this interval. The initial profile 
has a single maximum $u_m$ at some point $x_m$ (see Fig.~\ref{fig1}(a)). 

After wave breaking moment a DSW appears and in the Gurevich-Pitaevskii approach the 
wave number $k(x,t)$ of the locally periodic modulated wave is considered in Whitham 
approximation as one of the modulation variables,
so that $k/(2\pi)$ presents ``a density of waves'' within the DSW. Hence the number of
waves spanned by DSW is equal to
\begin{equation}\label{eq2-1}
  N_{\text{DSW}}(t)=\frac1{2\pi}\int_{x_L(t)}^{x_R(t)} k(x,t)dx,
\end{equation}
where we assume that $x_L(t)$ and $x_R(t)$ denote the coordinates of the
small-amplitude and soliton edges, correspondingly. Evolution of $k(x,t)$ obeys the number
of waves conservation law (Ref.~\onlinecite{whitham})
\begin{equation}\label{eq2-2}
  \frac{\prt k}{\prt t}+\frac{\prt\om}{\prt x}=0,
\end{equation}
where $\om=kV$ is the frequency of the periodic travelling wave solution, $V$ is its phase
velocity. The coordinate $x_R(t)$ corresponds to the position of the leading soliton whose
motion along a smooth background does not change $N_{\text{DSW}}(t)$. On the contrary, at the
small-amplitude edge the wave number $k(x,t)$ is not equal to zero and here we have a
flux $\om$ of waves into the DSW region. 

Our starting point is an important remark made by Gurevich and Pitaevskii in
Ref.~\onlinecite{gp-87} that since the small-amplitude edge of the DSW propagates
with the group velocity
\begin{equation}\label{eq3-4}
  v_g=\frac{\prt\om}{\prt k}
\end{equation}
of the wave at this edge, where $\om=\om(u,k)$ is the wave frequency of a linear
wave propagating along the background with the amplitude $u$, differs from the
phase velocity
\begin{equation}\label{eq3-5}
  V(u,k)=\frac{\om(u,k)}k
\end{equation}
of a linear wave, then the length of DSW increases at this edge by $(v_g-V)dt$
in the time interval $dt$, so that the number of waves inside DSW increases
with time as
\begin{equation}\label{eq3-6}
  \frac{dN_{\text{DSW}}}{dt}=\frac1{2\pi}k(v_g-V)=
  \frac1{2\pi}\left(k\frac{\prt\om}{\prt k}-\om\right).
\end{equation}
Up to the sign, this expression can be regarded as a Doppler-shifted
frequency representing the flux of waves into the DSW region. 
If we integrate the above formula upon time from the wave breaking moment to
$t=+\infty$, then we get the following formula for the number of solitons
(see Ref.~\onlinecite{kamch-20b})
\begin{equation}\label{eq3-7}
  N=\frac1{2\pi}\int_0^{\infty}k(v_g-V)dt
  =\frac1{2\pi}\int_0^{\infty}\left(k\frac{\prt\om}{\prt k}-\om\right)dt.
\end{equation}
All the parameters in the integrand are to be calculated at the
small-amplitude edge $x_L(t)$ of the DSW at the moment $t$ of its evolution.

In Whitham's approximation, a typical wavelength inside a DSW is much smaller
than the size of the whole DSW and this corresponds to the quasi-classical
approximation of wave propagation. At the small-amplitude edge the wave is
linear and the well-known Hamilton's optico-mechanical analogy (see, e.g.,
Ref.~\onlinecite{lanczos}) can be applied to propagation of the wave packet moving along
the path of the small-amplitude edge. According to this analogy, the motion of
this edge can be interpreted as a motion of a classical particle with momentum
$k$ and Hamiltonian $\om(u,k)$. Then the integrand in (\ref{eq3-7}) is interpreted
as a Lagrangian of this classical particle and the integral is equal to the action
$S$ produced by such a particle during its motion:
\begin{equation}\label{eq3-8}
  N=\frac{S}{2\pi}.
\end{equation}

Thus, our task is to develop the Hamilton theory of propagation of the small-amplitude
edge, extend it to the soliton edge, and to calculate asymptotic number of solitons
with the use of Eq.~(\ref{eq3-7}).

\section{General theory}

Now we take into account that the dependence of the Hamiltonian $\om(u,k)$
on the coordinate $x$ of the particle is carried on via the dependence of the
background simple wave $u(x,t)$ along which the small-amplitude short wavelength
perturbation of DSW propagates at this edge. Evolution of $u(x,t)$, on the
contrary to the short wavelength propagation of the wave packet perturbation,
is determined by the dispersionless hydrodynamic approximation of simple-wave type,
so that $u(x,t)$ obeys the equation (\ref{eq8-5}).
%\begin{equation}\label{eq4-9}
%  \frac{\prt u}{\prt t}+V_0(u)\frac{\prt u}{\prt x}=0,
%\end{equation}
%where
%\begin{equation}\label{eq4-10}
%  V_0(u)=V(u,0)
%\end{equation}
%is the velocity of unidirectional propagation of long waves along the background $u$
%which results in evolution of the profile $u(x,t)$ with time according to Eq.~(\ref{eq4-9}).
This Hopf equation for the simple-wave evolution can be easily solved for a given
initial distribution $u_0(x)$ (see, e.g., Ref.~\onlinecite{Whitham-74}),
\begin{equation}\label{eq4-11}
  x-V_0(u)t=\ox(u),
\end{equation}
where $\ox(u)$ is the function inverse to the initial distribution $u=u_0(x)$.
If we consider initial pulses in the form of a localized hump (see Fig.~1(a))
then the inverse function consists of two branches $\ox_1(u)$ and $\ox_2(u)$
(see Fig.~1(b)) and Eq.~(\ref{eq4-11}) determines in an implicit form the 
dependence $u=u(x,t)$ for each branch.

The specific dependence of the Hamiltonian $\om(u(x,t),k)$ on $x$ and $t$ via the
solution (\ref{eq4-11}) of Eq.~(\ref{eq8-5}) leads to important consequences. In particular,
the Hamilton equations
\begin{equation}\label{eq4-12}
  \frac{dx}{dt}=\frac{\prt\om}{\prt k},\qquad \frac{dk}{dt}=-\frac{\prt\om}{\prt x}
\end{equation}
together with Eq.~(\ref{eq8-5}) give at once
\begin{equation}\nonumber
  \frac{dk}{dt}=-\frac{\prt\om}{\prt u}\frac{\prt u}{\prt x},\quad
  \frac{du}{dt}=\frac{\prt u}{\prt x}\frac{dx}{dt}+\frac{\prt u}{\prt t}=
  -\left(V_0-\frac{\prt\om}{\prt k}\right)\frac{\prt u}{\prt x},
\end{equation}
and their ratio yields the equation
\begin{equation}\label{eq4-13}
  \frac{dk}{du}=\frac{\prt\om/\prt u}{V_0-\prt\om/\prt k}
\end{equation}
obtained by El in Ref.~\onlinecite{el-05}. Here the right-hand side depends only on 
$u$ and $k$, so its solution gives
\begin{equation}\label{eq4-14}
  k=k(u,q),
\end{equation}
where $q$ is the integration constant. The value $u=0$ corresponds to the initial 
moment of DSW formation when it shrinks to a point in the Gurevich-Pitaevskii
approach, that is the small-amplitude edge merges here with the soliton edge where
$k\to0$. This determines the boundary condition
\begin{equation}\label{eq4-15}
  k=0\quad\text{at}\quad u=0
\end{equation}
for Eq.~(\ref{eq4-13}) and specifies the value of the integration constant $h$
along the small-amplitude edge path.
After such a specification, the wave number $k=k(u)$ depends solely on $u$. 
As a result, if we consider the evolution of the pulse with a
step-like initial condition $u=u_0=\mathrm{const}$, then the solution $k=k(u)$
gives us the value $k(u_0)$ of the wave number at the small-amplitude edge
propagating along the constant background $u=u_0$ and, hence, the velocity
$v_g(k(u_0))$ of its propagation. This approach suggested by El, Ref.~\onlinecite{el-05},
permitted one to solve a number of interesting problems with step-like
initial conditions, 
Refs.~\onlinecite{el-05,egs-06,egkkk-07,ep-11,lh-13,hoefer-14,ckp-16,hek-17,ams-18}.

The solution (\ref{eq4-14}) satisfying the initial condition (\ref{eq4-15}) describes
the motion of the wave packet (its ray) at the small-amplitude DSW edge. Apparently, 
the Hamilton equations (\ref{eq4-12}) have more general character and describe the 
motion of wave packets along the background $u=u(x,t)$ with arbitrary initial conditions.
Since along each ray found in this way we have $q=\mathrm{const}$, then the variable
\begin{equation}\label{eq5-17}
  q=q(u,k)
\end{equation}
defined implicitly by Eq.~(\ref{eq4-14}) must satisfy the equation $q_t+v_gq_x=0$.
Combining this equation with Eq.~(\ref{eq8-5}), we arrive at the system
\begin{equation}\label{eq5-18}
  \frac{\prt u}{\prt t}+V_0(u)\frac{\prt u}{\prt x}=0,\quad 
  \frac{\prt q}{\prt t}+v_g(u)\frac{\prt q}{\prt x}=0
\end{equation}
with characteristic velocities equal to the limiting Whitham velocities at the 
small-amplitude edge. Thus, Eqs.~(\ref{eq5-18}) comprise continuation of the
Whitham equations on a smooth part of the pulse and at the small-amplitude edge 
of the DSW the variable $q$ can be regarded as a Riemann invariant of the
Whitham equations in this limit. Obviously, the system (\ref{eq5-18}) has a very
general nature and it often arises in description of the problem of 
interaction of linear wave packets with mean flow (see, e.g. 
Ref.~\onlinecite{ceh-19} and references within). It is worth noticing that in our 
approach the expression (\ref{eq5-17}) is obtained by means of solving 
Eq.~(\ref{eq4-13}) rather then by diagonalization of Eqs.~(\ref{eq8-5}) and 
(\ref{eq2-2}) although both methods are equivalent, of course.

Evidently, a similar reduction of the Whitham equations must exist at the soliton
edge of DSW and the question is how to find the Riemann invariant which corresponds
to the characteristic velocity $V_s$ equal to the speed of the leading soliton in DSW.
A hint to answering this question can be found in an old remark of Stokes fist
published in \S252 of the book Ref.~\onlinecite{lamb} and later reproduced in the form
of the letter to Lamb in Ref.~\onlinecite{stokes}. Stokes noticed that propagation
of the small amplitude soliton's tails (``outskirts'' according to his terminology) 
is governed by the same linearized equations that are used for description of propagation
of linear travelling waves, so that the expression $\exp[i(kx-\om t)]$ for the linear wave 
is replaced by the expression $\exp[-\tk(x-V_st)]$ for the tail at $x\to+\infty$. This
means that if we make the replacement $k\to i\tk$ in the dispersion law $\om(u,k)$
for linear waves and define
\begin{equation}\label{eq6-19}
  \tom(u,\tk)=-i\om(u,ik),
\end{equation}
then the soliton velocity is given by
\begin{equation}\label{eq6-20}
  V_s=\frac{\tom(u,\tk)}{\tk},
\end{equation}
where $\tk$ has the physical meaning of the inverse half-width of soliton.
This remark turned out to be very useful both in concrete studies of nonlinear wave
propagation (see, e.g., Refs.~\onlinecite{ai-77,dkn-03}) and in the theory of DSWs
for non-completely-integrable equations
(see Refs.~\onlinecite{el-05,egs-06,egkkk-07,ep-11,lh-13,hoefer-14,ckp-16,hek-17,ams-18}).

We assume that the same replacement transforms Eq.~(\ref{eq5-17}) into the
Riemann invariant
\begin{equation}\label{eq6-21}
  \ttq=\ttq(u,\tk)
\end{equation}
for the reduction
\begin{equation}\label{eq6-22}
  \frac{\prt u}{\prt t}+V_0(u)\frac{\prt u}{\prt x}=0,\quad
  \frac{\prt \ttq}{\prt t}+V_s\frac{\prt \ttq}{\prt x}=0
\end{equation}
of the Whitham equations at the soliton edge of DSW. Our assumption is
confirmed by checking its validity for the completely integrable equations
with known Whitham equations in the Riemann diagonal form (see Appendix A).
It is easy to check that
this transformation casts the Hamilton equations (\ref{eq4-12}) to the form 
\begin{equation}\label{eq6-23}
  \frac{dx}{dt}=\frac{\prt\tom}{\prt \tk},\qquad 
  \frac{d\tk}{dt}=-\frac{\prt\tom}{\prt x},
\end{equation}
and again these equations together with the first Eq.~(\ref{eq6-22}) yield
\begin{equation}\label{eq6-24}
  \frac{d\tk}{du}=\frac{\prt\tom/\prt u}{V_0-\prt\tom/\prt \tk}.
\end{equation}
Under certain assumptions, this equation was derived by El, Ref.~\onlinecite{el-05},
from the number of waves conservation law (\ref{eq2-2}).

By construction, the invariant $\ttq(u,\tk)$ is constant along trajectories
defined as solutions of Eqs.~(\ref{eq6-23}), so these trajectories can be regarded 
as paths of solitons with fixed values of $\ttq$. However, $\ttq$ changes along
the path $x_R(t)$ of the soliton edge determined by the solution of the equation
\begin{equation}\label{eq6-24b}
  \frac{dx_R}{dt}=V_s=\frac{\tom(u,\tk)}{\tk}.
\end{equation}
Apparently, this leading soliton path should be an envelope of paths of solitons 
with fixed values of $\ttq$. In a sense, at each moment of time the leading soliton 
is represented by an instant location of some soliton having invariant $\ttq$ when 
it touches the curve representing the path of the soliton edge of DSW. In fact,
this mechanism of edge formation as envelopes functions applies to the general
form of DSW appearing after wave breaking including its small-amplitude edge.
In such general situations the Whitham system does not reduce to one (for step-like
initial conditions) or two (for quasi-simple DSWs propagating into the quiescent 
medium) equations and we do not know beforehand the value of
the corresponding edge Riemann invariant $q$ or $\ttq$.  This
qualitative picture of DSW evolution agrees with known particular solutions
of Whitham equations for the KdV case which describe evolution of shocks after 
wave breaking of quadratic and cubic initial profiles (see Appendix B).

So far an initial distribution was assumed to be quite arbitrary. Now we turn to
the case of localized initial pulse shown in Fig.~\ref{fig1}(a),
so that it evolves into $N$ solitons.
To estimate the integral in Eq.~(\ref{eq3-7}) and to find $N$, we need to trace the 
variation of $u$ with time $t$ at the small-amplitude edge for the general form 
of the initial simple-wave pulse. This can be achieved by means of the following 
reasoning (see Refs.~\onlinecite{kamch-19,kamch-20}).
The small-amplitude edge propagates with the group velocity (\ref{eq3-4}), that is
during the time interval $dt$ it moves to the distance $dx=v_gdt$. Since this path
lies on the surface $u=u(x,t)$ of the dispersionless solution, the relation $dx/dt=v_g$
must be compatible with Eq.~(\ref{eq4-11}) representing this surface. For
a parametric representation $t=t(u)$ and $x=x(u)$ of the small-amplitude path, the
differentiation of (\ref{eq4-11}) with respect to $u$ and elimination of $dx/dt=v_g$
yields the equation
\begin{equation}\label{eq5-16}
  [v_g(u)-V_0(u)]\frac{dt}{du}-V_0'(u)t=\ox'(u).
\end{equation}
This linear differential equation $t(u)$ can be easily solved with the initial condition
$t=0$ at $u=0$ what gives us the dependence $t_1(u)$ of time $t$ on $u$ for the period of 
evolution when the small-amplitude edge propagates along the first branch of the dispersionless
solution corresponding to $\ox_1$. After the moment when $t$ reaches the time $t_1(u_m)$,
we have to solve Eq.~(\ref{eq5-16}) with the initial condition $t=t_1(u_m)$ at $u=u_m$
and this gives us the dependence $t_2(u)$ corresponding to propagation of the
small-amplitude edge along the second branch of the dispersionless solution. 
As a result, we obtain the function $t=t(u)$ for the total process of the pulse evolution
and this function together with the already
known functions $k(u)$ and $\om(u,k(u))$ permit us to calculate the number of
solitons with the use of Eq.~(\ref{eq3-7}). In concrete situations such a 
calculation can often be done without much difficulty and we shall illustrate the
method by a simple example in the next section.

\section{Example}

We consider here formation of solitons from a pulse $u_0(x)$ whose evolution is governed
by the generalized KdV equation
\begin{equation}\label{eq15-v1}
  u_t+V_0(u)u_x+u_{xxx}=0,
\end{equation}
which under certain conditions for $V_0(u)$, $V(0)=0$, has periodic and soliton
solutions (see Ref.~\onlinecite{el-05}). Linearization of this equation yields the
dispersion law of linear waves 
\begin{equation}\label{eq16-v1}
  \om(u,k)=V_0(u)k-k^3,
\end{equation}
so that Eq.~(\ref{eq4-13}) reduces to
$$
3k\frac{dk}{du}=V_0'(u),
$$
and its solution with the boundary condition Eq.~(\ref{eq4-15}) has the form
(see Ref.~\onlinecite{el-05})
\begin{equation}\label{eq17-v1}
  k(u)=\sqrt{\frac23V_0(u)}.
\end{equation}
Consequently, the group velocity at the small-amplitude edge propagating along
background with the amplitude $u$ is equal to $v_g(u)=-V_0(u)$. Then Eq.~(\ref{eq5-16}) 
becomes 
\begin{equation}\label{eq19-v1}
  -2V_0(u)\frac{dt}{du}-V_0'(u)t=\ox'(u)
\end{equation}
and for two branches shown in Fig.~\ref{fig1}(b) its solution reads 
(see Ref.~\onlinecite{kamch-19})
\begin{equation}\label{eq20-v1}
\begin{split}
  t(u)=t_1(u)=&-\frac1{2\sqrt{V(u)}}\int_0^u\frac{\ox_1'(u_1)}{\sqrt{V(u_1)}}\,du_1,\\
  t(u)=t_2(u)=&-\frac1{2\sqrt{V(u)}}\Bigg\{\int_0^{u_m}\frac{\ox'_1(u)}{\sqrt{V(u_1)}}\,du_1\\
  &+\int_{u_m}^u\frac{\ox'_2(u)}{\sqrt{V(u_1)}}\,du_1\Bigg\}.
  \end{split}
\end{equation}
Substitution of these expressions into Eq.~(\ref{eq3-7}) leads after simple transformations
to the formula
\begin{equation}\label{eq21-v1}
\begin{split}
  N=\frac{(2/3)^{3/2}}{2\pi}&\Bigg\{\int_0^{u_m}du\frac{V_0'(u)}2\int_u^{u_m}
  \frac{(\ox_2'-\ox_1')du_1}{\sqrt{V_0(u_1)}}\\
 & +\int_0^{u_m}\sqrt{V_0(u)}(\ox_2'-\ox_1')du\Bigg\}.
  \end{split}
\end{equation}
Here the double integral reduces to the ordinary one by means of evident integration
by parts with account of $V_0(0)=0$, so that we get the final expression 
\begin{equation}\label{eq22-v1}
\begin{split}
  N&=\frac1{2\pi}\int_0^{u_m}\sqrt{\frac23V_0(u)}\,(\ox_2'-\ox_1')\,du\\
  &=\frac1{2\pi}\int_{-l}^{0}\sqrt{\frac23V_0(u_0(x))}\,\,dx.
  \end{split}
\end{equation}
Remembering the formula (\ref{eq17-v1}) for the wave number, we obtain 
\begin{equation}\label{eq23-v1}
  N=\frac1{2\pi}\int k[u_0(x)]dx.
\end{equation}

The expression Eq.~(\ref{eq23-v1}) agrees with asymptotic formulas for the number of
solitons known for completely integrable equations and similar calculations for some 
other non-completely integrable equations lead to the final 
expressions which can also written in the form (\ref{eq23-v1}) what indicates its generality.
The general proof of this expression was suggested in Refs.~\onlinecite{egkkk-07,egs-08}
on the basis of extension of the notion of the DSW wave number $k$ beyond the DSW
region. In the next Section we present modification of this proof which clarifies
some its important points.

\section{Formula for the number of solitons}

\begin{figure}
\includegraphics[width=8cm]{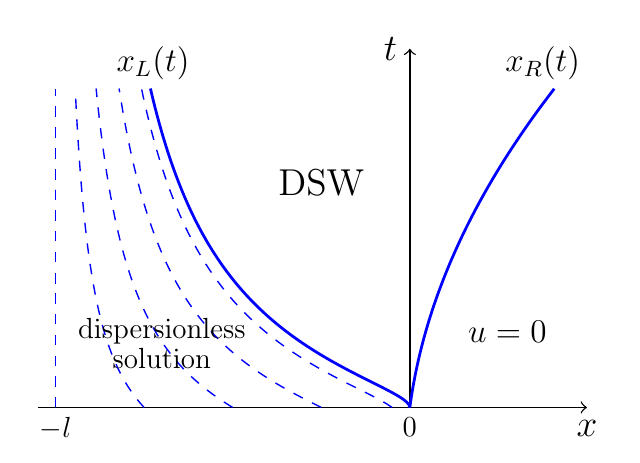}
\caption{Three regions distinguished in the waves structure evolves
from the initial profile $u_0(x)$: dispersionless solution for $x<x_L(t)$,
DSW for $x_L(t)\leq x\leq x_R(t)$, quiescent medium for $x>x_R(t)$. 
Dashed lines describe paths of wave
packets propagating along the dispersionless solution and forming the
distribution of $k_1(x,t)$.}
\label{fig2}
\end{figure}

The pulse evolved from the initial distribution depicted in Fig.~\ref{fig1}
consists in the Gurevich-Pitaevskii approximation from three parts: on the right
of the soliton edge $x_R(t)$ we have the quiescent medium with $u=0$, the DSW is
located between the two edges $x_L(t)\leq x\leq x_R(t)$, and on the left of the
small-amplitude edge $x_L(t)$ we have the dispersionless solution (see Fig.~\ref{fig2}).
The smooth evolution of the pulse outside DSW obeys Eq.~(\ref{eq8-5}) whose solution
$u(x,t)$ is given in implicit form by Eqs.~(\ref{eq4-11}) for two branches $\ox_1(u)$ 
and $\ox_2(u)$. Linear wave packets can propagate along this smooth background and
paths of these packets are given by solutions of the Hamilton equations (\ref{eq4-12})
for certain choice of initial conditions. We know that at the small-amplitude edge
the limiting Riemann invariant $q$ of the Whitham equations can be defined and its
value here is equal to $q=0$. This equality can be regarded as an expression of the
Gurevich-Meshcherkin assumption (see Ref.~\onlinecite{gm-84}) that in quasi-simple
DSWs the value of preserved dispersionless Riemann invariant is transferred through
a DSW.  This means that we can define an additional Riemann
invariant $q$ in the smooth region as an extension of one of the limiting Riemann 
invariants $u,q$ of Whitham equations to the whole dispersionless region. 
Thus, in the dispersionless region we have $q=0$ and, consequently, after 
substitution of this value into Eq.~(\ref{eq4-14}), we obtain the extension of
the function $k(u)=k(u,0)$ to the smooth region which yields distribution 
$k=k_1(x,t)=k[u(x,t)]$ of wave numbers as an extension of DSW's wave number 
at the small-amplitude edge to the whole smooth region (see Fig.~\ref{fig2}). One can 
say that this is a specific property of the Whitham approximation: although the 
amplitude of oscillations is equal here zero in this approximation, the notion of the
wavelength of waves, entering into the DSW region, still has physical meaning.
Obviously, $k_1(x,t)=0$ for $x<-l$ since here $u=0$ and $h=0$. Solutions of the equation
\begin{equation}\label{eq7-25}
  \frac{dx}{dt}=\left.\frac{\prt\om}{\prt k}\right|_{k=k_1(x,t)}
\end{equation}
with the initial condition $x(0)=x_0$, $-l\leq x_0\leq0$, give us a family of rays
along which wave packets propagate when they are radiated from points $x=x_0$ with the
carrying wave numbers $k[u_0(x_0)]$; they are shown by dashed lines in Fig.~\ref{fig2}. 
In particular, the ray radiated from the wave
breaking point $x_0=0$ gives us the path of the small-amplitude edge of the DSW.
We denote this path as $x=x_L(t)$.

Now we define the number of waves $N_{\text{smooth}}(t)$ corresponding to the defined above
distribution $k_1(x,t)$:
\begin{equation}\label{eq7-26}
\begin{split}
  N_{\text{smooth}}&=\frac1{2\pi}\int_{-l}^{x_L(t)}k_1(x_1,t)dx_1\\
  &=\frac1{2\pi}\int_{-l}^{x_L(t)}k[u(x_1,t)]dx_1.
  \end{split}
\end{equation}
It complements the number of waves $N_{\text{DSW}}$ entered into the DSW region up 
to the moment of time $t$ (see Eq.~(\ref{eq2-1})). The number $N_{\text{DSW}}$ changes 
with time according to Eq.~(\ref{eq3-6}), so let us calculate the derivative of 
$N_{\text{smooth}}$ with respect to $t$:
\begin{equation}\nonumber
  \frac{dN_{\text{smooth}}}{dt}=\frac1{2\pi}\left\{\frac{dx_L}{dt}k_1(x_a(t),t)+
  \int_{-l}^{x_L(t)}\frac{\prt k_1(x_1,t)}{\prt t}dx_1\right\}.
\end{equation}
We can substitute Eq.~(\ref{eq7-25}) with $x=x_a(t)$ into the first term. Then,
by definition the function $k_1(x,t)$ satisfies the number of waves conservation
law (\ref{eq2-2}) and this statement can easily be checked with the help of Eqs.~(\ref{eq8-5})
and (\ref{eq4-13}), so in the second term the integrand ${\prt k_1(x_1,t)}/{\prt t}$
can be replaced by $-{\prt \om(x_1,t)}{\prt x_1}$, and after integration we get
\begin{equation}\label{eq7-27}
  \frac{d N_{\text{smooth}}}{dt}=
  \frac1{2\pi}\left(k\frac{\prt\om}{\prt k}-\om\right)_{x=x_L(t)}.
\end{equation}
This is equal to Eq.~(\ref{eq3-6}) with opposite sign, that is 
$N_{\text{smooth}}+N_{\text{DSW}}=\mathrm{const}$. At last, since in the limit
$t\to\infty$ we have $N_{\text{DSW}}\to N$, $N_{\text{smooth}}\to0$ and for $t\to0$
we have $N_{\text{DSW}}\to 0$, 
$N_{\text{smooth}}\to (1/(2\pi))\int_{-\infty}^{\infty}k[u(x,0)]dx$ and $u(x,0)=u_0(x)$, 
we arrive at the final formula for the number of solitons
\begin{equation}\label{eq7-28}
  N=\frac1{2\pi}\int_{-\infty}^{\infty}k[u_0(x)]dx,
\end{equation}
where the function $k(u)$ is the solution of Eq.~(\ref{eq4-13}) with the boundary 
condition Eq.~(\ref{eq4-15}). The presented here proof of Eq.~(\ref{eq7-28}) provides 
an explicit construction of the function $k_1(x,t)$ for wave numbers in the smooth 
region of the pulse introduced earlier in Refs.~\onlinecite{egkkk-07,egs-08}.
The formula (\ref{eq7-28}) was confirmed by numerical solutions of nonlinear wave equations
and it agrees very well with the results of recent experiments presented in 
Ref.~\onlinecite{maiden-20}.

Under some additional assumptions, the asymptotic distribution of solitons parameters
was obtained in Ref.~\onlinecite{egs-08}.

\section{Conclusion}

The notion of {\it quasi-simple DSWs} was first introduced in Ref.~\onlinecite{gkm-89}
for the KdV equation case as the shocks with only two Riemann invariants changing along
them. In this paper, we generalized this notion to nonlinear wave situations with
wave breaking of simple waves, so that, according to Gurevich-Meshcherkin conjecture,
one dispersionless Riemann invariant has the same value at both edges of the DSW
under consideration. This definition is not limited to the class of completely integrable
equations and is applicable to any nonlinear wave equations admitting propagation of 
solitons. As follows from the Gurevich-Pitaevskii remark made in Ref.~\onlinecite{gp-87} 
on the number of waves entering into the DSW region in a unit of time, propagation of
the high-frequency wave packet at the small-amplitude edge of DSW satisfies the Hamilton 
equations with the linear dispersion law playing the role of the Hamiltonian, 
Ref.~\onlinecite{kamch-20b}. This system of Hamilton equations is coupled
with the Hopf equation for evolution of the background field what results in the
diagonal form of the Whitham equations at the DSW edges.

Long ago G.~G.~Stokes remarked in Refs.~\onlinecite{stokes,lamb} that the expression
for soliton's velocity can be obtained from the dispersion law for linear waves
because the tails of a soliton obey the same linearized equations as the small-amplitude 
travelling  waves. We generalize here this observation to the symmetry relationships 
between equations at the small-amplitude and soliton edges:
\begin{eqnarray}
    k\quad & \Leftrightarrow &\quad \tk, \\
  \om(u,k) \quad & \Leftrightarrow& \quad \tom(u,\tk)=-i\om(u,i\tk), \\
  \frac{dk}{du}=\frac{\prt\om/\prt u}{V_0-\prt\om/\prt k}\quad &\Leftrightarrow&
\quad \frac{d\tk}{du}=\frac{\prt\tom/\prt u}{V_0-\prt\tom/\prt \tk}, \\
   v_g=\frac{\prt\om}{\prt k}\quad &\Leftrightarrow& \quad V_s=\frac{\tom(\tk)}{\tk}, \\
  q(u,k) \quad &\Leftrightarrow& \quad \ttq(u,\tk)=q(u,i\tk), \\
  \frac{\prt q}{\prt t}+v_g\frac{\prt q}{\prt x}=0 \quad &\Leftrightarrow& \quad
\frac{\prt \ttq}{\prt t}+V_s\frac{\prt \ttq}{\prt x}=0 , 
\label{eq-f}
\end{eqnarray}
$k$ is the wave number at the small-amplitude edge, $\tk$ is in the soliton's
inverse half-width at the soliton edge, and at both edges the background field
obeys the same equation
\begin{equation}\label{eq6b-10}
  \frac{\prt u}{\prt t}+V_0\frac{\prt u}{\prt x}=0.
\end{equation}
Equations (\ref{eq-f}) and (\ref{eq6b-10}) comprise the limiting Whitham 
equations at the DSW edges for the Riemann invariants $(u,q)$ or $(u,\ttq)$, respectively.

Generally speaking, paths of small-amplitude and soliton edges of DSW are represented 
by envelopes of characteristics of Whitham equations and their finding is not an
easy task in non-completely-integrable case. Important exceptions are the
situations with initial step-like distributions when velocities of both edges
can be found (see Ref.~\onlinecite{el-05}) and a quasi-simple DSW propagating into
a quiescent medium when the path of one its edge can be calculated and the
asymptotic velocity of the other edge can be found in the case of localized
pulses (see Refs.~\onlinecite{kamch-19,kamch-20}).

In case of localized quasi-simple DSW propagating into a quiescent medium
and evolving into a train of solitons, the asymptotic formula for their number
can be derived with the use of Gurevich-Pitaevskii theorem on the number of
oscillations entering into the DSW region. The resulting formulas agree with
the expression derived in Refs.~\onlinecite{egkkk-07,egs-08}.

At last, although we considered here a concrete problem of evolution of
quasi-simple DSWs, some results can be applied to other problems of 
interaction of linear modulated waves with mean flow; see, e.g., Ref.~~\onlinecite{ceh-19}.

To sum up, the presented in this paper theory unifies the previously obtained 
result into a consistent approach applicable to a wide class of quasi-simple DSWs.

\begin{acknowledgments}
I am grateful to G.~A.~El, N.~Pavloff and L.~P.~Pitaevskii for useful discussions.
This study was funded by RFBR, project number 20-01-00063.
\end{acknowledgments}

\appendix
\section{Limiting Whitham equations: KdV equation case}

We shall consider the KdV equation 
\begin{equation}\label{eq7-29}
  u_t+6uu_x+u_{xxx}=0
\end{equation}
for which the Whitham system modulation equations can be written in diagonal form
with Riemann invariants $r_1,r_2,r_3$. Near the small-amplitude edge a DSW is 
approximated by the small-amplitude solution  (see, e.g., Ref.~\onlinecite{kamch-20b})
\begin{equation}\label{eq7-30}
\begin{split}
  &u(x,t)=r_3+(r_2-r_1)\cos[2\sqrt{r_3-r_1}\,(x-Vt)],\\
  & V=2(2r_1+r_3), \quad r_2-r_1\ll r_3-r_1,
  \end{split}
\end{equation}
where the Riemann invariants $r_1,r_3$  obey the limiting Whitham equations
\begin{equation}\label{eq7-31}
  \frac{\prt r_1}{\prt t}+(12r_1-6r_3)\frac{\prt r_1}{\prt x}=0,\quad
  \frac{\prt r_3}{\prt t}+6r_3\frac{\prt r_3}{\prt x}=0.
\end{equation}
As follows from Eq.~(\ref{eq7-30}), the background field and the wave number are
expressed in terms of $r_1,r_3$ by the formulas
\begin{equation}\label{eq7-32}
  u=r_3,\qquad k=2\sqrt{r_3-r_1}.
\end{equation}
Hence, we get $r_1=u-k^2/4=q$ and Eqs.~(\ref{eq7-31}) take the form
\begin{equation}\label{eq7-31}
  \frac{\prt q}{\prt t}+(6u-3k^2)\frac{\prt q}{\prt x}=0,\quad
  \frac{\prt u}{\prt t}+6u\frac{\prt u}{\prt x}=0,
\end{equation}
which coincides with the system Eqs.~(\ref{eq5-18}) with account of the dispersion law
\begin{equation}\label{eq7-34}
  \om(u,k)=6uk-k^3,\quad v_g=\frac{\prt\om}{\prt k}=6u-3k^2,\quad V_0(u)=6u
\end{equation}
of linear waves for a linearized Eq.~(\ref{eq7-29}).

Now, at the opposite edge of DSW the leading soliton has the form
\begin{equation}\label{eq7-35}
\begin{split}
  & u(x,t)=r_1+\frac{2(r_3-r_1)}{\cosh^2[\sqrt{r_3-r_1}(x-V_st)]}, \\
  & V_s=2(r_1+2r_3),
  \end{split}
\end{equation}
and the Whitham equations reduce to
\begin{equation}\label{eq7-36}
  \frac{\prt r_1}{\prt t}+6r_1\frac{\prt r_1}{\prt x}=0,\quad
  \frac{\prt r_3}{\prt t}+(2r_1+4r_3)\frac{\prt r_3}{\prt x}=0.
\end{equation}
We get expressions for the background field and the inverse half-width of soliton
from Eq.~(\ref{eq7-35}),
\begin{equation}\label{eq7-37}
  u=r_1,\qquad \tk=2\sqrt{r_3-r_1},
\end{equation}
so $r_3=r_1+\tk^2/4=\ttq$ and Eqs,~(\ref{eq7-36}) take the form
\begin{equation}\label{eq7-38}
 \frac{\prt u}{\prt t}+6u\frac{\prt u}{\prt x}=0,\quad
 \frac{\prt \ttq}{\prt t}+(6u+\tk^2)\frac{\prt \ttq}{\prt x}=0,
\end{equation}
coinciding with Eqs.~(\ref{eq6-22}) with account of
\begin{equation}\label{eq7-39}
  \tom(u,k)=6u\tk+\tk^3,\quad V_s=\frac{\tom}{ \tk}=6u+\tk^2,\quad V_0(u)=6u.
\end{equation}

Similar symmetry between equations for the DSW's edges can be proved 
for other completely integrable equations.

\section{Paths of DSW edges as envelopes}

First we consider situation with wave breaking of a parabolic pulse with
$u_0(x)=\sqrt{-x}, x\leq0,$ which belongs to the quasi-simple type. 
In the KdV equation theory, at the soliton edge with $r_1=u=0,r_3=\ttq$ 
the Whitham system Eqs.~(\ref{eq7-36}) reduces to
\begin{equation}\label{239-1}
  \frac{\prt\ttq}{\prt t}+4\ttq\frac{\prt\ttq}{\prt x}=0
\end{equation}
and its global solution becomes (see Ref.~\onlinecite{gkm-89,kamch-20b})
\begin{equation}\label{239-3}
  x-4\ttq t=-\frac8{15}\ttq^2.
\end{equation}
These are the characteristic curves near the soliton edge and for their
envelope the differentiation of Eq.~(\ref{239-3}) with respect to $\ttq$ gives
the relation $\ttq=(15/4)t$. Then $V_s=\tk^2=4\ttq=15t$ and the path of the
soliton edge is given by
\begin{equation}\label{239-3b}
  x_R=\frac{15}2t^2
\end{equation}
in agreement with the known result (see Ref.~\onlinecite{gkm-89,kamch-20b}).

Now we turn to a more complicated situation with the generic Gurevich-Pitaevskii 
problem on wave breaking of a cubic initial profile with $u_0(x)=(-x)^{1/3}$. 
In this case all three Riemann invariants are changing within the DSW 
and the global solution of Whitham equations was
found in Ref.~\onlinecite{potemin} (see also Ref.~\onlinecite{kamch-20b}).
At the small-amplitude edge it reduces to
\begin{equation}\nonumber
  x -(12r_1-6r_3)t=\frac{1}{5}(-16r_1^3+8r_1^2r_3+2r_1r_3^2+r_3^3).
\end{equation}
It was shown that at this edge we have $r_3=u$, $r_1=-u/4$, so the above equation
becomes 
\begin{equation}\label{239-5}
  x+9u=\frac14u^3
\end{equation}
and along envelope of these curves we get $u=\sqrt{12t}$. Then the group velocity
is equal to $v_g=-18\sqrt{3}\,t^{1/2}$ and integration of $dx_L/dt=v_g$ yields
for the small-amplitude path the formula
\begin{equation}\label{239-7}
  x_L=-12\sqrt{3}\,t^{3/2}
\end{equation}
in agreement with Refs.~\onlinecite{potemin,kamch-20b}.

In a similar way at the soliton edge the global solution gives
\begin{equation}\nonumber
  x-(2r_1+4r_3)t=-\frac{1}{35}(5r_1^3+6r_1^2r_3+8r_1r_3^2+16r_3^3),
\end{equation}
and it was shown that here $r_1=u$, $r_3=-3u/4$, hence this equation becomes
\begin{equation}\label{239-8}
  x+ut=\frac1{20}u^3.
\end{equation}
Along envelopes of these curves we get $u=-\sqrt{20t/3}$, $V_s=-u$, so
integration of $dx_R/dt=V_s$ yields
\begin{equation}\label{239-7}
  x_R=\frac49\sqrt{15}\,t^{3/2}
\end{equation}
again in agreement with Refs.~\onlinecite{potemin,kamch-20b}.

These examples demonstrate essential difference between step-like initial conditions,
simple wave pulse propagating into a quiescent medium, and the general simple 
wave initial pulse.

\end{document}